\documentclass[]{emulateapj}
\usepackage{graphics,wrapfig,amsmath,color,rotating,longtable}
\usepackage{ wasysym, url }

\begin{document}

\title{2011 HM$_{102}$: Discovery of a High-Inclination L5 Neptune Trojan \\In the Search for a post-Pluto {\it New Horizons} Target}
\author{ Alex H. Parker$^{1\dagger}$,  Marc W. Buie$^{2}$, David J. Osip$^{3}$, Stephen D.J. Gwyn$^{4}$, Matthew J. Holman$^{1}$, David M. Borncamp$^{2}$, John R. Spencer$^{2}$, Susan D. Benecchi$^{5}$, Richard P. Binzel$^{6}$, Francesca E. DeMeo$^{6}$, S\'{e}bastian Fabbro$^{4}$, Cesar I. Fuentes$^{7}$, Pamela L. Gay$^{8}$, JJ Kavelaars$^{4}$, Brian A. McLeod$^{1}$, Jean-Marc Petit$^{9}$, Scott S. Sheppard$^{5}$, S. Alan Stern$^{2}$, David J. Tholen$^{10}$, David E. Trilling$^{7}$, Darin A. Ragozzine$^{11,1}$, Lawrence H. Wasserman$^{12}$, and the Ice Hunters$^{13}$.
}
\email{$\dagger$ aparker@cfa.harvard.edu}
\affil{\emph{ $^{1}$Harvard-Smithsonian Center for Astrophysics, $^{2}$Southwest Research Institute, $^{3}$Carnegie Observatories, Las Campanas Observatory, $^{4}$Canadian Astronomy Data Centre, National Research Council of Canada, $^{5}$Carnegie Institute of Washington, Department of Terrestrial Magnetism, $^{6}$Massachusetts Institute of Technology, $^{7}$Northern Arizona University, $^{8}$Southern Illinois University, $^{9}$Universite de Franche Comte, CNRS, UTINAM, $^{10}$University of Hawaii, Institute for Astronomy, $^{11}$University of Florida, $^{12}$Lowell Observatory, and $^{13}$see appendix for list of contributing Ice Hunters. } }
\shortauthors{Parker et al.}

\begin{abstract}
We present the discovery of a long-term stable L5 (trailing) Neptune Trojan in data acquired to search for candidate Trans-Neptunian objects for the New Horizons spacecraft to fly by during an extended post-Pluto mission. This Neptune Trojan, 2011 HM$_{102}$, has the highest inclination ($29.4^\circ$) of any known member of this population. It is intrinsically brighter than any single L5 Jupiter Trojan at $H_V\sim8.18$. We have determined its $gri$ colors (a first for any L5 Neptune Trojan), which we find to be similar to the moderately red colors of the L4 Neptune Trojans, suggesting similar surface properties for members of both Trojan clouds.  We also present colors derived from archival data for two L4 Neptune Trojans (2006 RJ$_{103}$ and  2007 VL$_{305}$), better refining the overall color distribution of the population. In this document we describe the discovery circumstances, our physical characterization of 2011 HM$_{102}$, and this object's implications for the Neptune Trojan population overall. Finally, we discuss the prospects for detecting 2011 HM$_{102}$ from the New Horizons spacecraft during their close approach in mid- to late-2013.

\end{abstract}


\maketitle

\clearpage

\section{Introduction}

The New Horizons spacecraft will encounter Pluto on July 14$^{\mbox{th}}$ 2015 (Stern, 2008), after which, if an extended mission is approved, the spacecraft will alter course using onboard fuel reserves to target a more distant, much smaller Trans-Neptunian Object (TNO). This post-Pluto encounter will likely be the only opportunity for a close flyby with any spacecraft of a member of this distant population of minor planets in the foreseeable future. No currently known object is accessible with the spacecraft's estimated post-Pluto impulse budget of approximately $\Delta v \sim 120$ m s$^{-1}$, so a coordinated survey effort is ongoing in order to identify and characterize potential targets (Buie 2012). This survey targets the sky position of objects which will fall into the accessible path of the New Horizons spacecraft. In 2011-2012 the search fields are very near Neptune's trailing triangular Lagrange point (L5). Sheppard \& Trujillo (2010a) discovered the first L5 Neptune Trojan, 2008 LC$_{18}$. Our survey for a post-Pluto New Horizons target has serendipitously yielded the discovery of 2011 HM$_{102}$, a second long-term stable Neptune Trojan in the L5 cloud. 

To date, eight other long-term stable Neptune Trojans are known;\footnote{See list maintained at \\ \url{http://www.minorplanetcenter.net/iau/lists/NeptuneTrojans.html}} seven of these occupy the leading L4 cloud, and one occupies the trailing L5 cloud. There are other objects known to be temporarily co-orbital with Neptune; for example, the Minor Planet Center lists 2004 KV$_{18}$ as an L5 Neptune Trojan, but numerical integrations demonstrate that it becomes unstable on very short timescales, indicating that it is likely a scattering TNO or Centaur which is currently undergoing a transient resonant period, and does not represent a primordial Neptune Trojan (Gladman et al. 2012, Horner \& Lykawka 2012, Guan et al. 2012). The eight stable Neptune Trojans occupy a broad range of inclinations ($1.3^\circ - 28.1^\circ$) and libration amplitudes ($16^\circ-50^\circ$ peak-to-peak, Lykawka et al. 2009). 

The new L5 Neptune Trojan identified by our survey is not only stably resonant over Gyr timescales, it is also more highly inclined than any other known stable Neptune Trojan ($i=29.4^\circ$) and brighter than any other known stable L5 Trojan object in the Solar System ($H_V\simeq8.18$). It may be a candidate for long-range imaging from the New Horizons spacecraft in late 2013, when it flies by at a minimum distance of approximately 1.2 AU. Observations from the spacecraft may constrain the phase behavior of the surface at large phase angles, and provide a validation exercise for future long-range imaging of other TNOs in the post-Pluto mission phase.

In this letter we outline the discovery circumstances of 2011 HM$_{102}$, its current state of physical and orbital characterization, and prospects for a New Horizons long-range observational campaign.

\section{Discovery of 2011 HM$_{102}$}

A coordinated observational campaign has been undertaken to identify a post-Pluto encounter target for New Horizons. Between 2004 and 2005, a wide survey of moderate depth was performed with the Subaru telescope and the SuprimeCam imager intended to search for bright candidates. In 2011 the survey strategy was modified to target the now much smaller core of the sky-plane distribution of accessible objects (given the current understanding of the orbital distribution of the Kuiper Belt, eg. Petit et al. 2011) with much deeper observations in order to leverage the steep luminosity function of TNOs. Magellan MegaCam and IMACS, Subaru Suprime-Cam, and CFHT MegaPrime were all brought to bear on the field in a coordinated effort. 

The fields of interest are deep in the galactic plane, near Galactic coordinates $l \sim 12^\circ$, $b \sim -5^\circ$ (Ecliptic $\lambda \sim 277^\circ$, $\beta \sim 2^\circ$, Buie 2012), and are extremely crowded with background stars. Detection of faint moving sources requires high image quality observations and the application of difference imaging to reduce source confusion. Three independent difference imaging pipelines were developed to process the data in parallel, ensuring recovery of as large a fraction of moving objects in the fields as possible. All observations were referenced to a single master astrometric catalog, produced by the CFHT MegaPrime camera and tied to 2MASS astrometry using the MegaPipe calibration software (Gwyn 2008). This catalog went much fainter than 2MASS, allowing us to utilize long exposure times while still having many non-saturated stars to generate high-precision WCS solutions, and increased the precision and accuracy of our WCS solutions over all imagers we used compared to what would have been possible with the 2MASS stars alone.

Because the multiple independent pipelines were applied in some cases to the same data, not all astrometric measurements are entirely independent. See the appendices for a table of astrometry and an extended explanation of our approach to treating astrometric records from multiple pipelines.

In 2011, the target search area was approximately 2 square degrees, though this area was co-moving with the typical rates of TNOs accessible to New Horizons so the actual sky area is somewhat larger. CFHT, Subaru and Magellan observations covered subsets of this region to varying depth, typically at least $r\sim25$ and at the deepest reaching a 50\% completeness depth of $r\sim26.1$ in the central 0.18 square degree field (this depth was measured by implanting synthetic moving sources of known brightness into our data, then blindly recovering them with our pipelines simultaneously with real TNOs). At the current state of reduction, 25 probable new TNOs have been discovered in the 2011 data (as of the submission of this paper).

2011 HM$_{102}$ was first detected in Magellan IMACS observations from July 1---2, 2011, while at a heliocentric distance of $\sim27.8$ AU.  Recoveries in observations from Subaru, Magellan, and CFHT spanning April 29, 2011 to August 30, 2011 indicated that 2011 HM$_{102}$'s semi-major axis was consistent with that of Neptune, but with large uncertainty. During the first re-observation of the New Horizons fields in 2012, early-evening observations were made of 2011 HM$_{102}$ (which is now significantly outside the nominal New Horizons survey fields) and the one-year arc confirmed that the object was in a low-eccentricity orbit with a semi-major axis within $0.1\%$ of Neptune's, consistent with the orbit of an object in 1:1 mean-motion resonance. All astrometry is available in Table A1.

2011 HM$_{102}$ was recovered in a parallel effort by the citizen scientists involved in Ice Hunters\footnote{\url{http://www.icehunters.org/}} in the data acquired at Magellan over April 29---30 2011. The Ice Hunters users provided a list of transients for one of our pipelines by identifying sources in difference images by visual inspection. Users that contributed to the recovery are listed in the Appendix. 

2011 HM$_{102}$ is roughly a magnitude brighter ($r\sim22.55$) than the next brightest TNO discovered.  To date, no other object has been found in any of the New Horizons fields which is consistent with a Neptune Trojan, even though our survey has reached a depth at least 2.5 magnitudes fainter (from $r\sim25$ up to $26.1$) over the rest of the survey field. For populations of minor planets with power-law luminosity functions of moderate slope ($\alpha \sim 0.4-1.2$) which are continuous over the dynamic range of a given survey, the peak detection rate is generally found near the survey's 50\% completeness magnitude.  Adopting a Heaviside-function efficiency curve truncating at $r=25$, for any continuous power-law luminosity function with slope $\alpha$ steeper than $\sim0.52$, we would expect $>95$\% of any Neptune Trojans detected by our survey to be fainter than 2011 HM$_{102}$. However, if the luminosity function truncates at $r\sim23.5$, as indicated by Sheppard \& Trujillo (2010b), then much steeper bright-object slopes would be consistent with the detection of a single object with the luminosity of 2011 HM$_{102}$ in our survey. We therefore interpret the detection of 2011 HM$_{102}$ along with the lack of any fainter detections as further evidence for a lack of intermediate-size planetesimals as suggested by Sheppard \& Trujillo (2010b).

\section{Orbital and Physical Properties of 2011 HM$_{102}$}

The current barycentric orbital elements for 2011 HM$_{102}$ are listed in Table 1. Of note is its very high inclination of $29.4^\circ$, making it the highest inclination Neptune Trojan known. The only other known long-term stable L5 Neptune Trojan, 2008 LC$_{18}$, also has a high inclination at $27.6^\circ$. The New Horizons survey fields fall approximately $2^\circ$ from the Ecliptic, and so are significantly more sensitive to lower inclination objects. 2011 HM$_{102}$ was discovered at an ecliptic latitude of $\sim 2.46^\circ$, and given its inclination it spends less than 9\% of its time at lower latitudes. 

The discovery of such a high inclination Trojan prior to the discovery of any low inclination Trojans in near-ecliptic fields lends support to the L5 Neptune Trojans being a highly excited population, as suggested by Sheppard \& Trujillo (2010a). For comparison, $\sim$10\% of all Jupiter Trojans with $H<10$ have higher inclinations than 2011 HM$_{102}$.

\subsection{Resonant behavior}\label{res_beh}

Using the \textit{mercury6} N-body integrator (Chambers, 1999), we integrated the orbits of 7,000 clones of 2011 HM$_{102}$ for $10^7$ years into the future to verify that within its uncertainties it is a long-term resonant object. These clones were sampled from a multivariate normal prior in the cartesian orbital basis, centered on the best-fit orbit with the covariance matrix generated by the \textit{fit\_radec} and \textit{abg\_to\_xyz} routines developed in association with Bernstein \& Kushalani (2000), assuming uniform astrometric uncertainties of $0.1"$ on all data points. After translating their orbits into the heliocentric basis of \textit{mercury6}, the clones were integrated with the $hybrid$ integration algorithm, along with the giant planets (with the mass of the Sun augmented by the mass of the terrestrial planets). All clones librated stably for the entire $10^7$ year integration. Using the lower 3-sigma Wilson score interval (Wilson 1927) we find that 2011 HM$_{102}$ is resonant over $>10^7$ years with $>99.9$\% confidence. A DES-style three-clone, $10^7$ year integration and classification (Elliot et al. 2005) also indicates stable Trojan behavior over this timescale.

The integrated clones show average libration amplitudes of $a=19.4^\circ\pm0.8^\circ$ (peak-to-peak, measured over individual libration cycles and then averaged) over the $10^7$ year integration, with libration periods of $9545\pm 4$ years. The maximum libration amplitude over a single libration period experienced by any clone in the entire 10$^7$ years of integration was 26.5$^\circ$, well inside the limits of long-term stability ($60^\circ-70^\circ$, Nesvorn\'{y} \& Dones 2002). The approach of averaging individual libration cycle amplitudes removes the effects of longer-period oscillations in the orbit; defining the libration amplitude by computing the RMS of the resonant argument around its mean and assuming sinusoidal behavior results in an similar value of $a=19.2^\circ \pm 0.8^\circ$, while defining the amplitude as the difference between the largest and smallest resonant argument over the entire integration results in a larger estimate of $a=24.2^\circ \pm 0.8^\circ$. Regardless of definition, all of these estimates are comfortably within the limits of stability. The mean libration center was found to be $59.95^\circ \pm 0.08^\circ$, consistent with the ideal center of the L5 cloud. 

A subset of 100 clones (drawn from the same orbital distribution as the larger sample) were integrated for 1 Gyr. Again, no clone became unstable over the integration. Accounting for the smaller sample size, and again adopting the lower 3-sigma Wilson score interval, we can assert that 2011 HM$_{102}$ is stably resonant over a large fraction of the age of the Solar System with $> 92\%$ confidence. The maximum libration amplitude reached over a single libration cycle by any of these 100 clones in the entire 1 Gyr of integration was 37.1$^\circ$.

\subsection{Colors of 2011 HM$_{102}$, 2007 VL$_{305}$, and 2006 RJ$_{103}$}

\begin{figure}[t]
\begin{centering}
\includegraphics[width=\columnwidth]{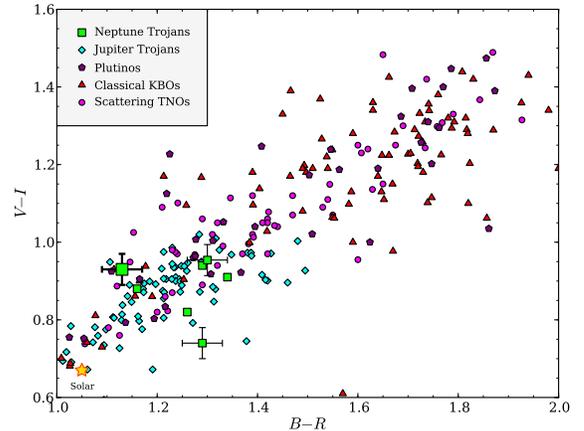}
\caption{ Photometric color indices of 2011 HM$_{102}$ (large green square with error bars) compared to other outer Solar System populations. Colors of other Neptune Trojans taken from Sheppard \& Trujillo (2006), while all other colors are compiled from the MBOSS database (Hainaut \& Delsanti 2002). Other Neptune Trojans marked with error bars are colors of 2007 VL$_{305}$ and 2006 RJ$_{103}$ measured in this work.}
\end{centering}
\end{figure}

On the night of UT May 23---24, 2012, observations with Magellan IMACS (f4 camera, 2$\times$4 array of 2k$\times$4k E2V CCDs, $1\times1$ binning, in fast read mode) measured the $gri$ colors of 2011 HM$_{102}$. Observations were made in photometric conditions, and the Landolt standard 107 351 (Landolt 1992: $V-R = 0.351 $, slightly bluer than known Neptune Trojans $\langle V-R \rangle \simeq 0.46$, Sheppard \& Trujillo 2006) was observed over the same airmass range ($z\sim1.2 - 2.2$) to determine photometric zero-points and extinction coefficients (no color term corrections were calculated or applied). Exposure times for 2011 HM$_{102}$ were 300s in each filter, and filters were cycled four times throughout the observations to ensure that any rotational lightcurve (though no evidence was seen for one) would not affect the color measurements. By choosing a standard star with similar color indices to our target population, we minimize potential differences between the SDSS photometric system and the native system of the IMACS camera and filters due to un-characterized color terms.

Due to field crowding, photometry was performed on difference images, thus the reduction of the target was unavoidably different from the reduction of the (stationary) standard star. The images in each band were PSF-matched (using the ISIS package, Alard \& Lupton 1998) to the image in that band taken at lowest airmass (to limit the propagation of extinction correction error from our photometric standard calibration). A template image was constructed by taking the minimum pixel value across the stack of the four PSF-matched images in each band, which was then subtracted off of each PSF-matched image. The motion of the target was sufficient that in the four visits in each filter, the PSF of the source did not overlap in all images in any pixel --- but the motion was not so great as to cause any significant elongation of the target's PSF. At this point, since all images are photometrically scaled to the image taken at lowest airmass, we estimated the zero-points and extinction coefficients for all difference images based on the lowest observed airmass in that band. Photometry was extracted with a fixed aperture radius of $\sim1$ FWHM, and while aperture corrections were determined for each image independently (from the PSF-matched, yet un-subtracted images), all were effectively identical (as expected from the PSF-matching).

The Solar System Object Search (SSOS) service (Gwyn, Hill \& Kavelaars 2012) provided by the Canadian Astronomy Data Centre also turned up un-published color measurements of two bright L4 Neptune Trojans --- 2007 VL$_{305}$, and 2006 RJ10. These data were collected by the CFHT MegaPrime camera in photometric conditions in November of 2010 through $gri$ filters. The MegaPipe software (Gwyn 2008) was used to perform photometric calibration for these data, and photometry for the two Trojans was extracted from the calibrated images. The measured color indices were then transformed into the SDSS system\footnote{Transformations for the MegaPrime camera and its filter system into the SDSS system can be found here: \\\url{http://www3.cadc-ccda.hia-iha.nrc-cnrc.gc.ca/megapipe/}}.

Table 1 lists the $gri$ colors of all three objects, as well as these colors translated into the \textit{BVRI} system using the transformations from Smith (2002). These transformations were verified as accurate for similar TNO surfaces by Sheppard \& Trujillo (2006). 

We find that the colors of 2011 HM$_{102}$ are moderately red; entirely consistent with the L4 Neptune Trojans measured by Sheppard \& Trujillo (2006) as well as those measured here. In general the colors of the Neptune Trojans are consistent with the neutral Centaurs or the Jupiter Trojans ($\langle V-R \rangle \simeq 0.445$, Fornasier et al. 2007); Figure 1 demonstrates the color of 2011 HM$_{102}$, the L4 Neptune Trojans, and their relation to other populations. The similar colors of the L4 Trojans and 2011 HM$_{102}$ supports both the L4 and L5 Neptune Trojan clouds sharing a common composition and environmental history, suggestive of a common origin. However, as demonstrated by the Jupiter Trojans' NIR spectral dichotomy, optical colors alone are not sufficient to rule out compositional variation in an otherwise apparently homogeneous population (Emery et al. 2011).

\begin{table*}[t]
\centering
\begin{tabular}{  lllll  }
\multicolumn{4}{c}{\bf Table 1: Orbit and Color Properties}\\
\hline
\multicolumn{4}{c}{2011 HM$_{102}$ J2000 Osculating Barycentric Orbital Elements at Epoch 2455680.8}\\
Semi-major axis & Eccentricity & Inclination & Lon. of Asc. Node \\
$30.109 \pm 0.002$ AU & $0.0803\pm0.0002$ & $29.3780^\circ \pm 0.0005^\circ$ & $100.9870^\circ \pm 0.0002^\circ$ \\
Argument of Periastron & JD of Periastron & Libration Amplitude$^a$ & Libration Period$^b$ \\
$\omega =152.2^\circ \pm 0.3^\circ$ & $2452464\pm46$ &$19.4^\circ\pm0.8^\circ$& $9545\pm 4$ years & \\
\hline
\multicolumn{4}{c}{Observed and Derived Magnitudes}\\
 2011 HM$_{102}$:   & $r=22.55\pm0.03^c$ & $V=22.75\pm0.04^c$  & $H_{\mbox{\tiny V(1,1,0)}} = 8.18^d$ &  \\
 2007 VL$_{305}$: & $r=22.76\pm0.03$ & $V=23.00\pm0.03$  & $H_{\mbox{\tiny V(1,1,0)}} = 8.5^d$ &  \\
 2006 RJ$_{103}$: & $r=22.03\pm0.02$ & $V=22.27\pm0.04$  & $H_{\mbox{\tiny V(1,1,0)}} = 7.4^d$ &  \\
\hline
\multicolumn{4}{c}{Observed and Derived Color Indices}\\
 2011 HM$_{102}$: & $g-r=0.51\pm0.04^c$ & $r-i=0.31\pm0.04^c$ & $g-i=0.82\pm0.04^c$ &  \\
 $B-V=0.72\pm0.04^c$ & $V-R=0.41\pm0.04^c$ & $R-I=0.52\pm0.04^c$& $B-I=1.66\pm0.04^c$\\
 2007 VL$_{305}$: & $g-r=0.62\pm0.05$ & $r-i=0.27\pm0.05$ & $g-i=0.89\pm0.05$ &  \\
 $B-V=0.83\pm0.05$ & $V-R=0.47\pm0.05$ & $R-I=0.48\pm0.05$& $B-I=1.78\pm0.05$\\
 2006 RJ$_{103}$: & $g-r=0.61\pm0.03$ & $r-i=0.06\pm0.04$ & $g-i=0.67\pm0.04$ &  \\
 $B-V=0.82\pm0.03$ & $V-R=0.47\pm0.03$ & $R-I=0.27\pm0.04$& $B-I=1.56\pm0.04$\\
\hline
\end{tabular}\\
{\footnotesize$^a$:  Mean peak-to-peak libration amplitude over $10^7$ year numerical integration. $^b$:  Mean  libration period over $10^7$ year numerical integration.  $^c$:  Uncertainties only represent estimate of shot noise, as sample was too small to empirically estimate scatter around the mean. $^d$:  Absolute magnitude at zero phase derived assuming phase slope of G=0.14 mag deg$^{-1}$ }\\
\end{table*}

\subsection{ Size and Population Comparisons }

We estimate a $V$-band absolute magnitude at zero phase of $H\simeq8.18$ for 2011 HM$_{102}$. Three other known L4 Neptune Trojans are brighter than this, but there is only one object in the Jupiter L5 cloud which is of similar brightness --- (617) Patroclus, $H\sim8.19$.  Interestingly, Patroclus is a binary system with components of nearly equal size (Merline et al. 2002, Marchis et al. 2006), indicating that if the albedos of the two populations are similar, there are no single L5 Jupiter Trojans physically larger than 2011 HM$_{102}$ --- unless, of course, 2011 HM$_{102}$ is \textit{also} a binary system. Adopting Patroclus' $V$-band albedo of $\sim0.045$ (Mueller et al. 2010), 2011 HM$_{102}$ is approximately 140 km in diameter.

%%% 

The sample of Jupiter Trojans with $H<10$ is essentially complete, and we estimate that approximately 5\% of this population falls within the latitude band that our survey covered at any given time. Assuming a similar inclination distribution for the Neptune Trojans, detecting one object in this latitude band would indicate a total population of approximately 20 objects. The orbit of 2011 HM$_{102}$ spends less than 2.7\% of its time within the same band, which would indicate a larger total population of $\sim37$ objects on similar orbits for every one object inside the latitude band of our survey. Given that our survey covered only a tiny fraction of the longitudinal extent of the L5 cloud, any estimate of the population based only on latitude coverage represent a lower limit on the population.  This indicates that the Neptune Trojan L5 cloud has at \textit{least} an order of magnitude more large ($d \gtrsim 100$ km) objects as Jupiter's L5 cloud. The results of Sheppard \& Trujillo (2010a) indicate similar populations size for the Neptune L4 and L5 Trojans, and so our detection compares favorably with the estimates by Sheppard \& Trujillo (2006) of large Neptune L4 Trojans being $5-20$ times more populous than L4 Jupiter Trojans of comparable size.

\begin{figure}[t]
\begin{centering}
\includegraphics[width=\columnwidth]{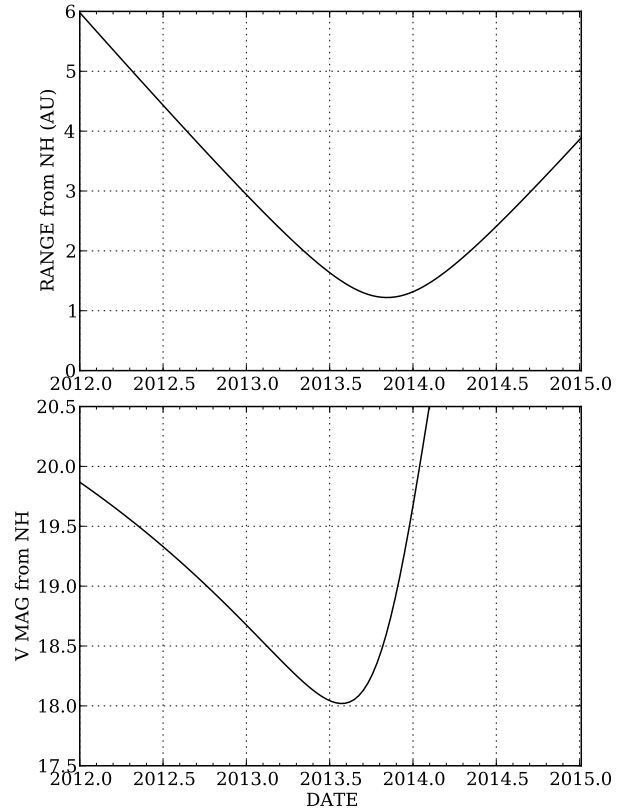}
\caption{ Range and apparent $V$-band magnitude of 2011 HM$_{102}$ as viewed from the New Horizons trajectory over the next 2.5 years. Magnitude assumes a phase function with G=0.14. Peak brightness occurs at a phase angle of approximately $45^\circ$.}
\end{centering}
\end{figure}

\section{Prospects for Long-Range Reconnaissance from New Horizons}

The New Horizons spacecraft will make its closest approach to 2011 HM$_{102}$ in late 2013, at a minimum distance of approximately 1.2 AU. Assuming an (H,G) phase curve parameterization with phase slope parameter G$=0.14$ (similar to asteroid surfaces), 2011 HM$_{102}$ will reach peak brightness as viewed from the spacecraft several months prior to closest approach. Figure 2 illustrates the separation of New Horizons and 2011 HM$_{102}$ over 2012---2015, and the predicted apparent V-band magnitude of 2011 HM$_{102}$ as seen from New Horizons, accounting for the changing viewing geometry and estimated phase curve. At peak, we estimate that 2011 HM$_{102}$ will appear approximately $V \sim 18$ from New Horizons. A single 10s exposure from the \textit{LOng-Range Reconnaissance Imager} (LORRI, Cheng et al. 2008) in $4\times4$ binning mode would expect to detect a $V\sim18$ source with SNR$\sim$4; multiple such exposures can be combined to provide useful SNR. No observation of a $d\sim100$ km object at this range would be resolved by the LORRI PSF, though binary companions may be resolvable. Such an observation may provide the opportunity to verify the procedures for imaging TNOs during long-range flybys, as there will be several such encounters in the Kuiper Belt. In addition, a detection from the unique vantage point of the New Horizons spacecraft would verify the large phase-angle behavior of TNO surfaces (un-observable from the Earth), which would provide useful information for predicting the outcome of other long-range encounters, and for planning the optimal navigation strategy for targeting these future flybys.

\section{Summary}

2011 HM$_{102}$ represents a surprising discovery in a survey not directly designed for the detection of Neptune Trojans. Its brightness compared to the depth of the survey support the evidence for a break in the Neptune Trojan luminosity function, and its colors suggest that the L4 and L5 clouds share similar physical properties and history. Because of its highly excited orbit, it spends very little time near to the libration center of the L5 Trojans or to the ecliptic (where the survey was directed) suggesting a large unseen population of similar objects. It is likely larger than any L5 Jupiter Trojan, and from its detection we infer that $d\gtrsim100$ km L5 Neptune Trojans are at least an order of magnitude more populous than Jupiter's L5 Trojans of similar size. 

Numerical integration suggests that 2011 HM$_{102}$ is stably resonant over the age of the Solar System with libration amplitudes comparable to other stable Neptune Trojans.

During mid- to late-2013, 2011 HM$_{102}$ may be detectable by the LORRI instrument on the New Horizons spacecraft, potentially allowing measurements of the phase curve of a TNO surface at far greater solar elongation than has ever been possible. Such a measurement would be valuable for predicting the success of detecting other TNOs at long range in the post-Pluto mission.

\section{Acknowledgements}

This paper includes data gathered with the 6.5 meter Magellan Telescopes located at the Las Campanas Observatory, Chile, as well as data collected at Subaru Telescope, which is operated by the National Astronomical Observatory of Japan, and on observations obtained with MegaPrime/MegaCam, a joint project of CFHT and CEA/DAPNIA, at the Canada-France-Hawaii Telescope (CFHT) which is operated by the National Research Council (NRC) of Canada, the Institute National des Sciences de l'Univers of the Centre National de la Recherche Scientifique of France, and the University of Hawaii. This research used the facilities of the Canadian Astronomy Data Centre operated by the National Research Council of Canada with the support of the Canadian Space Agency.  Magellan time was acquired through the CfA, MIT, Carnegie, and Arizona TACS, and CFHT time was acquired through the Canada, France, and Hawaii TACs. 

This work was supported in part by a NASA Keck PI Data Award, administered by the NASA Exoplanet Science Institute. Some of the data presented herein, obtained at the Subaru telescope, was made possible through a time swap from the W. M. Keck Observatory allocated to the National Aeronautics and Space Administration through the agency's scientific partnership with the California Institute of Technology and the University of California. The Observatory was made possible by the generous financial support of the W. M. Keck Foundation.

The authors wish to recognize and acknowledge the very significant cultural role and reverence that the summit of Mauna Kea has always had within the indigenous Hawaiian community. We are most fortunate to have the opportunity to conduct observations from this mountain.

%{\color{red} [ Editorial Note: The following appendices will be included as an online supplement, but are included here in the draft document for editorial simplicity. ]}

\appendix

\section{ Ice Hunters }

A list of the users who contributed to the recovery of 2011 HM$_{102}$ through the Ice Hunters citizen science project:

{\sc  A. Agbedor, A. Assioli, E. Baeten, T. D. Beer, P. Bel, M. C. Blanaru,
 M. Bovingdon, P. Brayshaw, T. Brydon, D. Cameron, J. Campos, E. Conseil, M. Cotton,
 C. Cripps, A. Crouthamel, J. Dadesky, J. M. Dawey, T. Demko, L. Dinsdale,
 G. Dungworth, A. Duvall, A. Erena, R. Evans, P. Fitch,
 R. Frasier, R. Gagliano, B. Gilbert, A. Gillis, V. Gonano, F. Helk,
 F. Henriquez, M. Herrenbruck, J. Herridge, D. Herron, T. Hodge,
 S. Ivanchenko, M. Kelp, C. Kindel, J. Koopmans, H. Krawczyk, A. Lamperti,
 D. V. Lessen, S. Li, N. Macklem, M. H. Massuda, A. Maya, M. T. Mazzucato,
 K. McCoy, P. A. McDonald, R. Mideke, G. Mitchell, V. Mottino, D. O'Connor,
 M. Olga, N. N. Paklin, A. Pandey, C. Panek, E. R. Pearsall, K. Pidgley,
 S. Pogrebenko, B. Replogle, J. Riley, K. Roovers, C. Schlesinger, T. Sieben,
 P. D. Stewart, S. R. Taylor, J. Thebarge, H. Turner, R. H.B. Velthuis,
 P. Verdelis, E. Walravens, B. Way, B. Wyatt, A. Zane, M. Zehner,
 D. R. Zeigler}

\section{Astrometry of 2011 HM$_{102}$}

As the data acquired for this project was reduced by multiple pipelines, our astrometric record includes overlapping measurements. While these measurements are not wholly independent, they are sufficiently distinct that we include all measurements here. Distinctions include reductions via different astrometric basis functions, difference imaging pipelines, and duration of bins in time. Future submissions to the Minor Planet Center of other discoveries by our team will also include all measurements. In the attached table, measurements made by each independent pipeline are labeled as ``SWRI,'' ``CFA,'' or ``HIA.'' Similarly, the data submitted to the Minor Planet Center is distinguished by unique headers for each pipeline. Treating the non-independence of overlapping measurements should be done at the time of any new orbit fitting; relative weighting of observations may be changed by the addition of new data.

\begin{center}
\begin{longtable}{  ccccc  }
%\begin{tabular}{  ccccc  }
\multicolumn{5}{c}{\bf Table A1: Multi-source Astrometry of 2011 HM$_{102}$}\\[0.5ex]
$\langle JD \rangle$ & $\alpha$ (J2000) & $\delta$ (J2000) & Obscode & Astcode \\[0.5ex]
\endfirsthead

{{Continued from previous page}} \\
$\langle JD \rangle$ & $\alpha$ (J2000) & $\delta$ (J2000) & Obscode & Astcode \\[0.5ex]
\endhead

\hline \multicolumn{5}{r}{{Continued on next page}} \\ 
\endfoot
\hline \hline
\endlastfoot

2455680.80225 & 18:34:51.760 & -20:32:49.67 & 304 & CFA \\ [0.5ex]
2455680.80402 & 18:34:51.756 & -20:32:49.70 & 304 & CFA \\ [0.5ex]
2455680.80935 & 18:34:51.738 & -20:32:49.77 & 304 & CFA \\ [0.5ex]
2455680.81112 & 18:34:51.734 & -20:32:49.78 & 304 & CFA \\ [0.5ex]
2455680.81202 & 18:34:51.731 & -20:32:49.79 & 304 & SWRI \\ [0.5ex]
2455680.81292 & 18:34:51.728 & -20:32:49.79 & 304 & CFA \\ [0.5ex]
2455680.81470 & 18:34:51.723 & -20:32:49.81 & 304 & CFA \\ [0.5ex]
2455680.81648 & 18:34:51.718 & -20:32:49.83 & 304 & CFA \\ [0.5ex]
2455680.82003 & 18:34:51.707 & -20:32:49.86 & 304 & CFA \\ [0.5ex]
2455680.82182 & 18:34:51.701 & -20:32:49.89 & 304 & CFA \\ [0.5ex]
2455681.70947 & 18:34:49.081 & -20:32:58.60 & 304 & SWRI \\ [0.5ex]
2455681.83360 & 18:34:48.702 & -20:32:59.96 & 304 & CFA \\ [0.5ex]
2455681.83535 & 18:34:48.695 & -20:32:59.99 & 304 & CFA \\ [0.5ex]
2455681.83711 & 18:34:48.689 & -20:32:59.97 & 304 & CFA \\ [0.5ex]
2455681.83888 & 18:34:48.687 & -20:32:59.98 & 304 & CFA \\ [0.5ex]
2455681.84064 & 18:34:48.679 & -20:33:00.00 & 304 & CFA \\ [0.5ex]
2455681.84153 & 18:34:48.674 & -20:32:59.97 & 304 & SWRI \\ [0.5ex]
2455681.84242 & 18:34:48.673 & -20:32:59.98 & 304 & CFA \\ [0.5ex]
2455681.84418 & 18:34:48.667 & -20:33:00.01 & 304 & CFA \\ [0.5ex]
2455681.84594 & 18:34:48.661 & -20:33:00.06 & 304 & CFA \\ [0.5ex]
2455681.84772 & 18:34:48.654 & -20:33:00.03 & 304 & CFA \\ [0.5ex]
2455681.84947 & 18:34:48.652 & -20:33:00.00 & 304 & CFA \\ [0.5ex]
2455710.92902 & 18:32:28.899 & -20:39:01.45 & 568 & SWRI \\ [0.5ex]
2455711.07807 & 18:32:27.939 & -20:39:03.75 & 568 & SWRI \\ [0.5ex]
2455711.96730 & 18:32:22.319 & -20:39:16.57 & 568 & SWRI \\ [0.5ex]
2455712.91888 & 18:32:16.197 & -20:39:30.90 & 568 & SWRI \\ [0.5ex]
2455714.08665 & 18:32:08.564 & -20:39:48.51 & 568 & SWRI \\ [0.5ex]
2455743.58703 & 18:28:33.956 & -20:48:04.88 & 304 & CFA \\ [0.5ex]
2455743.58961 & 18:28:33.936 & -20:48:04.80 & 304 & CFA \\ [0.5ex]
2455743.59153 & 18:28:33.919 & -20:48:04.81 & 304 & CFA \\ [0.5ex]
2455743.59345 & 18:28:33.909 & -20:48:04.78 & 304 & CFA \\ [0.5ex]
2455743.59537 & 18:28:33.891 & -20:48:04.90 & 304 & CFA \\ [0.5ex]
2455743.59729 & 18:28:33.881 & -20:48:05.00 & 304 & CFA \\ [0.5ex]
2455743.59921 & 18:28:33.867 & -20:48:05.01 & 304 & CFA \\ [0.5ex]
2455743.60007 & 18:28:33.855 & -20:48:05.15 & 304 & SWRI \\ [0.5ex]
2455743.60007 & 18:28:33.875 & -20:48:05.05 & 304 & SWRI \\ [0.5ex]
2455743.60116 & 18:28:33.856 & -20:48:05.09 & 304 & CFA \\ [0.5ex]
2455743.60308 & 18:28:33.839 & -20:48:05.13 & 304 & CFA \\ [0.5ex]
2455743.60500 & 18:28:33.821 & -20:48:05.06 & 304 & CFA \\ [0.5ex]
2455743.60692 & 18:28:33.806 & -20:48:05.11 & 304 & CFA \\ [0.5ex]
2455743.67551 & 18:28:33.271 & -20:48:06.31 & 304 & CFA \\ [0.5ex]
2455743.67744 & 18:28:33.263 & -20:48:06.49 & 304 & CFA \\ [0.5ex]
2455743.67936 & 18:28:33.247 & -20:48:06.40 & 304 & CFA \\ [0.5ex]
2455743.68130 & 18:28:33.232 & -20:48:06.59 & 304 & CFA \\ [0.5ex]
2455743.68322 & 18:28:33.214 & -20:48:06.63 & 304 & CFA \\ [0.5ex]
2455743.68502 & 18:28:33.214 & -20:48:06.67 & 304 & SWRI \\ [0.5ex]
2455743.68502 & 18:28:33.234 & -20:48:06.63 & 304 & SWRI \\ [0.5ex]
2455743.68514 & 18:28:33.198 & -20:48:06.64 & 304 & CFA \\ [0.5ex]
2455743.68706 & 18:28:33.186 & -20:48:06.65 & 304 & CFA \\ [0.5ex]
2455743.68898 & 18:28:33.173 & -20:48:06.77 & 304 & CFA \\ [0.5ex]
2455743.69091 & 18:28:33.154 & -20:48:06.63 & 304 & CFA \\ [0.5ex]
2455743.69284 & 18:28:33.152 & -20:48:06.77 & 304 & CFA \\ [0.5ex]
2455743.69477 & 18:28:33.139 & -20:48:06.99 & 304 & CFA \\ [0.5ex]
2455743.76066 & 18:28:32.624 & -20:48:07.99 & 304 & CFA \\ [0.5ex]
2455743.76258 & 18:28:32.599 & -20:48:08.02 & 304 & CFA \\ [0.5ex]
2455743.76450 & 18:28:32.585 & -20:48:08.10 & 304 & CFA \\ [0.5ex]
2455743.76644 & 18:28:32.571 & -20:48:08.09 & 304 & CFA \\ [0.5ex]
2455743.76836 & 18:28:32.556 & -20:48:08.11 & 304 & CFA \\ [0.5ex]
2455743.77017 & 18:28:32.550 & -20:48:08.24 & 304 & SWRI \\ [0.5ex]
2455743.77017 & 18:28:32.554 & -20:48:08.21 & 304 & SWRI \\ [0.5ex]
2455743.77029 & 18:28:32.547 & -20:48:08.24 & 304 & CFA \\ [0.5ex]
2455743.77220 & 18:28:32.527 & -20:48:08.22 & 304 & CFA \\ [0.5ex]
2455743.77413 & 18:28:32.518 & -20:48:08.30 & 304 & CFA \\ [0.5ex]
2455743.77605 & 18:28:32.497 & -20:48:08.24 & 304 & CFA \\ [0.5ex]
2455743.77797 & 18:28:32.490 & -20:48:08.38 & 304 & CFA \\ [0.5ex]
2455743.77991 & 18:28:32.475 & -20:48:08.36 & 304 & CFA \\ [0.5ex]
2455743.84587 & 18:28:31.972 & -20:48:09.54 & 304 & CFA \\ [0.5ex]
2455743.84779 & 18:28:31.949 & -20:48:09.56 & 304 & CFA \\ [0.5ex]
2455743.84972 & 18:28:31.941 & -20:48:09.70 & 304 & CFA \\ [0.5ex]
2455743.85164 & 18:28:31.925 & -20:48:09.76 & 304 & CFA \\ [0.5ex]
2455743.85356 & 18:28:31.913 & -20:48:09.71 & 304 & CFA \\ [0.5ex]
2455743.85549 & 18:28:31.893 & -20:48:09.75 & 304 & CFA \\ [0.5ex]
2455743.85736 & 18:28:31.898 & -20:48:09.65 & 304 & SWRI \\ [0.5ex]
2455743.85741 & 18:28:31.878 & -20:48:09.77 & 304 & CFA \\ [0.5ex]
2455743.85935 & 18:28:31.867 & -20:48:09.81 & 304 & CFA \\ [0.5ex]
2455743.86127 & 18:28:31.848 & -20:48:09.83 & 304 & CFA \\ [0.5ex]
2455743.86319 & 18:28:31.848 & -20:48:09.85 & 304 & CFA \\ [0.5ex]
2455743.86510 & 18:28:31.819 & -20:48:09.93 & 304 & CFA \\ [0.5ex]
2455743.86736 & 18:28:31.806 & -20:48:09.98 & 304 & CFA \\ [0.5ex]
2455743.86936 & 18:28:31.789 & -20:48:09.94 & 304 & CFA \\ [0.5ex]
2455744.53049 & 18:28:26.803 & -20:48:22.00 & 304 & CFA \\ [0.5ex]
2455744.53241 & 18:28:26.787 & -20:48:22.04 & 304 & CFA \\ [0.5ex]
2455744.53434 & 18:28:26.771 & -20:48:22.16 & 304 & CFA \\ [0.5ex]
2455744.53626 & 18:28:26.760 & -20:48:22.15 & 304 & CFA \\ [0.5ex]
2455744.53818 & 18:28:26.744 & -20:48:22.20 & 304 & CFA \\ [0.5ex]
2455744.54000 & 18:28:26.729 & -20:48:22.28 & 304 & SWRI \\ [0.5ex]
2455744.54012 & 18:28:26.730 & -20:48:22.26 & 304 & CFA \\ [0.5ex]
2455744.54204 & 18:28:26.712 & -20:48:22.24 & 304 & CFA \\ [0.5ex]
2455744.54398 & 18:28:26.698 & -20:48:22.32 & 304 & CFA \\ [0.5ex]
2455744.54590 & 18:28:26.683 & -20:48:22.35 & 304 & CFA \\ [0.5ex]
2455744.54784 & 18:28:26.669 & -20:48:22.36 & 304 & CFA \\ [0.5ex]
2455744.54976 & 18:28:26.656 & -20:48:22.45 & 304 & CFA \\ [0.5ex]
2455744.63822 & 18:28:25.975 & -20:48:24.04 & 304 & CFA \\ [0.5ex]
2455744.64014 & 18:28:25.962 & -20:48:24.10 & 304 & CFA \\ [0.5ex]
2455744.64206 & 18:28:25.942 & -20:48:24.13 & 304 & CFA \\ [0.5ex]
2455744.64398 & 18:28:25.930 & -20:48:24.17 & 304 & CFA \\ [0.5ex]
2455744.64591 & 18:28:25.914 & -20:48:24.19 & 304 & CFA \\ [0.5ex]
2455744.64772 & 18:28:25.898 & -20:48:24.24 & 304 & SWRI \\ [0.5ex]
2455744.64784 & 18:28:25.899 & -20:48:24.25 & 304 & CFA \\ [0.5ex]
2455744.64977 & 18:28:25.885 & -20:48:24.26 & 304 & CFA \\ [0.5ex]
2455744.65169 & 18:28:25.866 & -20:48:24.34 & 304 & CFA \\ [0.5ex]
2455744.65362 & 18:28:25.852 & -20:48:24.34 & 304 & CFA \\ [0.5ex]
2455744.65554 & 18:28:25.838 & -20:48:24.37 & 304 & CFA \\ [0.5ex]
2455744.65747 & 18:28:25.822 & -20:48:24.41 & 304 & CFA \\ [0.5ex]
2455744.74847 & 18:28:25.123 & -20:48:26.05 & 304 & CFA \\ [0.5ex]
2455744.75039 & 18:28:25.109 & -20:48:26.12 & 304 & CFA \\ [0.5ex]
2455744.75233 & 18:28:25.093 & -20:48:26.16 & 304 & CFA \\ [0.5ex]
2455744.75425 & 18:28:25.080 & -20:48:26.21 & 304 & CFA \\ [0.5ex]
2455744.75618 & 18:28:25.062 & -20:48:26.26 & 304 & CFA \\ [0.5ex]
2455744.75702 & 18:28:25.054 & -20:48:26.25 & 304 & SWRI \\ [0.5ex]
2455744.75809 & 18:28:25.047 & -20:48:26.25 & 304 & CFA \\ [0.5ex]
2455744.76002 & 18:28:25.033 & -20:48:26.27 & 304 & CFA \\ [0.5ex]
2455744.76194 & 18:28:25.022 & -20:48:26.34 & 304 & CFA \\ [0.5ex]
2455744.76387 & 18:28:25.004 & -20:48:26.36 & 304 & CFA \\ [0.5ex]
2455744.76580 & 18:28:24.988 & -20:48:26.37 & 304 & CFA \\ [0.5ex]
2455744.78855 & 18:28:24.841 & -20:48:26.93 & 568 & SWRI \\ [0.5ex]
2455744.83772 & 18:28:24.427 & -20:48:27.70 & 304 & CFA \\ [0.5ex]
2455744.84927 & 18:28:24.349 & -20:48:27.86 & 304 & CFA \\ [0.5ex]
2455744.84947 & 18:28:24.374 & -20:48:28.10 & 304 & SWRI \\ [0.5ex]
2455744.85040 & 18:28:24.366 & -20:48:28.11 & 568 & SWRI \\ [0.5ex]
2455744.85120 & 18:28:24.340 & -20:48:27.87 & 304 & CFA \\ [0.5ex]
2455744.85313 & 18:28:24.322 & -20:48:27.91 & 304 & CFA \\ [0.5ex]
2455744.85506 & 18:28:24.307 & -20:48:27.99 & 304 & CFA \\ [0.5ex]
2455744.85697 & 18:28:24.286 & -20:48:27.89 & 304 & CFA \\ [0.5ex]
2455744.85902 & 18:28:24.278 & -20:48:28.19 & 304 & CFA \\ [0.5ex]
2455744.91279 & 18:28:23.888 & -20:48:29.32 & 568 & SWRI \\ [0.5ex]
2455776.84707 & 18:24:41.561 & -20:58:34.05 & 568 & HIA \\ [0.5ex]
2455776.85027 & 18:24:41.543 & -20:58:34.16 & 568 & HIA \\ [0.5ex]
2455776.86949 & 18:24:41.418 & -20:58:34.51 & 568 & HIA \\ [0.5ex]
2455802.52299 & 18:22:46.742 & -21:06:38.89 & 304 & SWRI \\ [0.5ex]
2455802.58703 & 18:22:46.549 & -21:06:40.14 & 304 & SWRI \\ [0.5ex]
2455802.65098 & 18:22:46.362 & -21:06:41.31 & 304 & SWRI \\ [0.5ex]
2455803.54200 & 18:22:43.883 & -21:06:57.83 & 304 & SWRI \\ [0.5ex]
2455803.60496 & 18:22:43.693 & -21:06:58.92 & 304 & SWRI \\ [0.5ex]
2456034.78353 & 18:45:10.468 & -21:37:16.00 & 304 & CFA \\ [0.5ex]
2456034.78540 & 18:45:10.463 & -21:37:16.06 & 304 & CFA \\ [0.5ex]
2456034.78716 & 18:45:10.461 & -21:37:16.09 & 304 & CFA \\ [0.5ex]
2456034.78892 & 18:45:10.459 & -21:37:16.12 & 304 & CFA \\ [0.5ex]
2456034.79069 & 18:45:10.460 & -21:37:16.13 & 304 & CFA \\ [0.5ex]
2456034.79244 & 18:45:10.456 & -21:37:16.15 & 304 & CFA \\ [0.5ex]
2456034.79421 & 18:45:10.452 & -21:37:16.14 & 304 & CFA \\ [0.5ex]
2456034.79597 & 18:45:10.454 & -21:37:16.17 & 304 & CFA \\ [0.5ex]
2456034.79774 & 18:45:10.452 & -21:37:16.18 & 304 & CFA \\ [0.5ex]
2456034.79951 & 18:45:10.450 & -21:37:16.26 & 304 & CFA \\ [0.5ex]
2456034.80128 & 18:45:10.449 & -21:37:16.24 & 304 & CFA \\ [0.5ex]
2456034.80303 & 18:45:10.444 & -21:37:16.26 & 304 & CFA \\ [0.5ex]
2456034.80496 & 18:45:10.444 & -21:37:16.16 & 304 & CFA \\ [0.5ex]
2456034.80574 & 18:45:10.443 & -21:37:16.30 & 304 & CFA \\ [0.5ex]
2456034.80672 & 18:45:10.442 & -21:37:16.27 & 304 & CFA \\ [0.5ex]

\end{longtable}
\end{center}

\end{document}